\documentstyle[12pt,epsfig]{article}
\def\nn{\nonumber}
\def\ee{\end{equation}}
\def\be{\begin{equation}}
\def\eea{\end{eqnarray}}
\def\bea{\begin{eqnarray}}
\def\eeas{\end{eqnarray*}}
\def\beas{\begin{eqnarray*}}
 
\begin{document}
\begin{titlepage}
\begin{center}
{\large\bf A composite bosons minimal basis for the pairing Hamiltonian }
\vspace*{1cm}

{M.B.Barbaro$^1$, A.Molinari$^1$, 
F.Palumbo$^2$ and M.R.Quaglia$^1$}

\vspace*{1cm}
{\it 
$^1$ Dipartimento di Fisica Teorica dell'Universit\`a di Torino
and INFN Sezione di Torino, via P.Giuria 1, I--10125, Torino, Italy\\
$^2$ INFN - Laboratori Nazionali di Frascati,P.O.Box 13, 
I--00044 Frascati, Italy}

\vspace*{1.5cm}

{\bf ABSTRACT}\\

\begin{quotation}
{\it The pairing interaction among identical nucleons 
in a single-particle level is treated in the hamiltonian formalism
using even Grassmann variables. A minimal (irreducible) basis having a remarkable symmetry property is set up using composite, commuting variables with a finite index of nilpotency.
Eigenvalues and eigenfunctions of given energy, 
seniority and zero third component of the angular momentum of the pairing 
hamiltonian are then found.
The eigenvectors, which cannot be cast solely in terms of composite 
bosons with angular momentum zero and two, are
expanded in the minimal basis with coefficients
analytically expressed in terms of a generalized hypergeometric 
function.}
\end{quotation}
\end{center}


\vspace*{1.5cm}
\noindent
Corresponding author: Maria B. Barbaro,
Dipartimento di Fisica Teorica,
Universit\`a di Torino,
Via P. Giuria 1,
I-10125 Torino - Italy,
Tel: 39-11-6707212,
Fax: 39-11-6707214,
e-mail: barbaro@to.infn.it

\end{titlepage}

A difficulty inherent to nuclear physics relates to the number 
of nucleons in a nucleus which is not large enough to exploit the 
field theoretical 
techniques as successfully as in condensed matter physics, but, 
on the other hand, {\it is} large enough to require a high dimensionality for
the basis in which the physics of the nucleus is treated. 
 
The standard approach to overcome this obstacle reduces the 
size of the basis 
by selecting the {\it relevant degrees of freedom}, namely those 
appropriate for the description of the low-lying nuclear excitations, 
and then by introducing the associated variables, for example as done
in the Arima and Iachello model \cite{Iac}. 

Clearly one would like to devise a systematic procedure to carry out this
reduction starting  
from an underlying, more fundamental hamiltonian expressed in 
terms of fermionic degrees of freedom.

Recently we have indeed tried an approach \cite{Bar97} to this problem 
in the path integral framework by 
performing a non-linear change of variables in the Berezin integral
defining the partition function of a fermionic system, the new integration 
variables then representing {\it composite bosons} of appropriate quantum 
numbers. 

As a simple example, we have applied this method to
the problem of an even number $N$ of nucleons sitting in a 
single-particle level with angular momentum $j$ (third component $m$) and 
interacting through the well-known pairing hamiltonian $H_P$.
We have been able to express 
the pairing action and the ground state wave function with zero seniority
in terms of the composites and to recover
the familiar formula for the ground state energy of $H_P$ \cite{Bar97}.
Moreover through this study we learned about several useful properties of these composites which have been applied to set up a 
new perturbative expansion in QCD \cite{Pal98}-\cite{Pal99}. 

However, to solve the problem of the excited states of $H_P$ 
(those with nonzero seniority $v$) 
and to extend the approach to encompass the long range 
quadrupole-quadrupole force as well as the pairing interaction 
has proven to be a hard task to perform. 
Therefore in this letter we address the $v\ne 0$ problem without resorting to 
the change of variables in the Berezin integral. 
Instead we explore whether and how the Fock basis of $H_P$, set up with 
determinants 
of single-particle states, can be reduced to a minimal dimension and 
whether the {\it minimal basis} can be expressed in terms of composite bosons.
Indeed one might expect the physics of $H_P$, which only acts among pairs 
of nucleons coupled to angular momentum $J=J_z=0$, to be describable by only 
two composite bosons having the two constituent nucleons 
coupled to $J=0$ (the first) and to any allowed non-vanishing 
angular momentum (the second).  

Specifically, in carrying out the program of reducing the 
natural Fock basis of $H_P$ we shall search for a minimal (irreducible) 
basis, fully solving the pairing problem, using building blocks 
associated with composite, commuting variables having a finite index of nilpotency.
Hence these are not to be viewed as {\it bona fide}
bosons, being made up of fermions satisfying the Pauli principle; and this is why we shall refer to them as composite bosons.
Notably the dimensions of the basis will turn out to be fixed 
by the number $n= N/2$ of pairs present in the problem. 

To illustrate how our approach works, let us start by recalling
that in the quasi-spin scheme one 
introduces operators annihilating (creating) pairs of particles in orbits having appropriate time-reversal properties, i.e. of the type 
\be
{\hat B}_J= \sqrt{\frac{\Omega}{2}}
\sum_{m=-j}^j \langle j m, j -m |J0\rangle {\hat a}_{-m}{\hat a}_{m} 
\label{Aj}
\ee
(and the hermitian conjugate).
In the above $\langle j m, j -m |J0\rangle$ is the usual 
Clebsch Gordan coefficient and  $\Omega =(2j+1)/2$. 
In terms of (\ref{Aj}) the pairing hamiltonian reads
\be
{\hat H}_P = -g_P {\hat B}^{\dag}_0 {\hat B}_0 ~.
\label{H_pairing}
\ee
As mentioned above,
one might conjecture that the eigenstates of $H_P$ could be set up in 
terms 
of the operators ${\hat B}_0$ and ${\hat B}_J$ where the specific value of the
index $J(\ne 0)$ in the latter should be irrelevant, since any $J\ne 0$ 
corresponds to a broken pair. For example, in the spirit of the Arima 
and Iachello model, one could set $J=2$. 

However in the framework of the creation and annihilation operators the
commutator 

\be
\left[ \hat{a}_{-m} \hat{a}_{m}, \hat{a}^{\dag}_{ m'} 
\hat{a}^{\dag}_{-m'}\right]= \delta_{mm'}\left( 1- 
\hat{a}^{\dag}_{m'} \hat{a}_{m} - 
\hat{a}^{\dag}_{-m} \hat{a}_{-m'}\right)
\label{commutatore}~,
\ee
which includes, e.g. for $J=J'=0$, 
\be
\left[ {\hat B}_0, ~{\hat B}_0^{\dag} \right]= 
\Omega\left(1- \frac{{\hat n}}{\Omega}\right)~,
\label{commutatoreA}
\ee
is non-canonical, thus rendering non-trivial the task of finding the 
eigenstates of $H_P$ (${\hat n}$ is the fermion number operator). 

An approach that circumvents this difficulty is the hamiltonian
framework employing Grassmann variables \cite{Fad80}. Here 
one exploits the isomorphism between the Fock space $F$ generated by a set 
of $N$ fermionic creation operators 
$\hat{a}_1^{\dag},\cdots \hat{a}_{N}^{\dag}$ 
and the so-called ${\cal{G}}^+$ algebra built with the set of  
anticommuting objects $\lambda^*_1,...\lambda^*_{N}$ (the generators 
of the algebra). The isomorphism is defined by mapping the vectors
$\hat{a}^{\dag}_1...\hat{a}^{\dag}_j|0>$ 
onto the elements 
$\lambda^*_1...\lambda^*_j$. 
The image of a generic vector $|\Psi> \in F$ under this mapping will be
denoted by $\Psi(\lambda^*)$.
Next, to a linear operator in normal form, 
 
\begin{equation}
\hat{{\cal O}}=\sum_{i_1...i_k} \sum_{j_1...j_k} 
{\cal O}^{i_1...i_k,j_1...j_k}\hat{a}^{\dag}_{i_1}...\hat{a}^{\dag}_{i_k}
\hat{a}_{j_1}...\hat{a}_{j_k}~, 
\end{equation}
is associated the following function of the Grassmann variables 

\begin{equation}
{\cal{O}}(\lambda^*,\lambda )=\sum_{i_1,...i_k,j_1...j_k}
{\cal{O}}^{i_1,...i_k,j_1...j_k}
\lambda^*_{i_1}...\lambda^*_{i_k}\lambda_{j_1}...\lambda_{j_k} ,
\label{OGrassmann}
\end{equation}
which allows one to define the kernel

\begin{equation}
K_{{\cal{O}}}(\lambda^*,\lambda)={\cal{O}}(\lambda^*,\lambda) 
\mu_+(\lambda^*\lambda)~.
\end{equation}
Then the action of $\hat{\cal O}$ on a state $\Psi$ reads 

\begin{equation}
(\hat{\cal O} \Psi)(\lambda^*)= \int [d\lambda'^* d\lambda'] 
K_{\cal{O}}(\lambda^*,\lambda') \mu_-(\lambda'^*\lambda')
\Psi(\lambda'^*)\ .
\label{eq:Opsi}
\end{equation}
In the above
\begin{equation}
\mu_{\pm}(\lambda^*\lambda)=e^{\pm \sum_i \lambda^*_i \lambda_i}.
\end{equation}

In this framework the pairing hamiltonian is then readily 
expressed in terms of the {\it even, nilpotent, commuting Grassmann variables}
\be
\varphi_m= (-1)^{j-m}\lambda_{-m}\lambda_{m}
\label{phi}
\ee
according to

\be
H_P = -g_P B^*_0 B_0
\ \ \ \mbox{with}\ \ \ 
B_0= \sum_{m=1/2}^j \varphi_m \ .
\ee

Now, in seeking a reduction of the basis dimensionality, we shall be 
guided by the two main features of $H_P$, namely that it
\begin{enumerate}
\item
is expressed solely in terms of the $\varphi$,
\item
is invariant for any permutation of the $\varphi$.
\end{enumerate}

These properties suggest also expressing the vectors of the basis 
in terms of the variables (\ref{phi}). For this scope, the action of 
$H_P$ on the latter should be explored.

This can be accomplished 
along the lines illustrated in \cite{Bar97}, the result being
\be
H_P \psi(\varphi^*) = \int [d\eta^* d\eta] K_P (\varphi^*,\eta) 
e^{\sum{\eta^*\eta}}\psi(\eta^*) = E \psi(\varphi^*)
\label{eq_eigenvalue}
\ee
where the integral is over the even elements of the Grassmann algebra as 
defined in \cite{Bar97} and the kernel is 
\be
K_P(\varphi^*,\eta)= H_P(\varphi^*,\eta) 
e^{\sum{\varphi^*\eta}}~.
\ee
Hence, the above-mentioned difficulty associated with the 
non-canonical nature of the commutator (\ref{commutatore}) disappears. 

We then attempt 
to diagonalize the $H_P$ associated with $n= N/2$ pairs in a basis 
spanned by states represented as products of $n$ factors $\varphi^*$'s, 
namely of the type
\be
\varphi^*_{m_1} \cdots \varphi^*_{m_n}~.
\label{base}
\ee

Since the number of states here is clearly
${\Omega}\choose{n}$, the very large reduction
of the basis dimension entailed by the choice of the variables (\ref{phi}) is 
apparent: indeed in terms of fermionic degrees of freedom, 
the corresponding basis would have a dimension ${2\Omega}\choose{2n}$.

Moreover, because the variables $\lambda$ anticommute,
each vector of the basis (\ref{base}) is antisymmetric with respect to the exchange 
of any pair of fermions, and hence the above basis fulfills the Pauli principle.  

We now explore whether, for a given $\Omega$ and $n$, 
the eigenstates of $H_P$ can be cast in the form
\be
\psi (\varphi^*) = \sum_{m=1}
^{{\Omega}\choose{n}} 
\beta_m {[\varphi^*_{m_1}
\cdots \varphi^*_{m_n}]}_m~,
\label{psi_generale}
\ee
the index $m$ identifying the set $\{ m_1, m_2\cdots m_n\}$ and
the $\beta_m$ being complex coefficients.
 
In the simple situation where $n=1$, the states in (\ref{psi_generale}) 
are indeed eigenstates of the pairing hamiltonian. 
In fact the eigenvalue equation, with ${\cal E}= E/g_P$, is 
\be
\left( {\cal E} +1\right) \beta_{m} + \sum_{p(\ne m)=1/2}^{j} \beta_{p} =0
\ee
and in the basis in (\ref{base}) the operator
${\cal H}_P= -\frac{H_P}{g_P}+{\cal E}$ 
is represented by a matrix of dimension $\Omega$ filled by ones
but for the principal diagonal, whose elements are given by ${\cal E}$+1.
This matrix is invariant under any permutation of the 
index $m$ which labels its rows and columns. 
Moreover, of the $\Omega$ real roots of the associated characteristic equation 
$$\left( {\cal E}+ \Omega \right) {\cal E}^{\Omega-1} =0$$
only two are distinct, namely the lower 
$ {\cal E}_0(n=1) =- \Omega$  with multiplicity $\delta=1$ 
and the upper one
$ {\cal E}_2(n=1) =0$ with multiplicity $\delta=\Omega-1$. 
The associated orthogonal eigenvectors are
\be
\Psi_0(n=1) = \frac{1}{\sqrt{\Omega}}\sum_{m=1/2}^j \varphi^*_m
\label{psi0_n1}
\ee
and 
\be
\Psi_2(n=1)= {\cal N} \sum_{m=1/2}^{j-1}\beta_m \left[
\varphi^*_m - \varphi^*_j\right]
\label{psi2_n1}
\ee
(${\cal N}$ is a normalization factor). 
Notice that into the eigenvector in (\ref{psi2_n1}) enter 
$\Omega-1$ parameters (the $\beta_m$), which correspond to the degeneracy
of the eigenvalue.
The above eigenvalues agree with the quasi-spin formalism and 
correspond to the states with seniority $v=0$ and $v=2$ respectively, 
$v$ being an {\it even non-negative number} representing the number of unpaired
particles.  

Since the case with 
$n=\Omega -1$ is identical, but for a shift in energy, to the one with 
$n=1$ pairs and, by extension,
the case with $n$ pairs is equivalent to the one with $\Omega-n$ pairs,
in the following we shall confine ourselves 
to consider only $n\le \frac{\Omega}{2}$.

When $n=2$, it is not so trivial to prove that (\ref{psi_generale}) is an 
eigenfunction of the pairing hamiltonian for any $\Omega$, although
this turns out to be the case. For example when $\Omega=4$ 
the eigenvalue equation is 
\be
\left( {\cal E} +2\right) \beta_{mn} + \sum_{p(\ne n,m)=1/2}^{j}
\left( \beta_{pm} + \beta_{pn}\right) =0
\ee
and ${\cal H}_P$ is represented by the $6\times 6$ matrix whose elements are
${\cal E}$+2 on the principal diagonal, 0 on the secondary diagonal and 1 
elsewhere. The associated characteristic equation 
$$({\cal E}+6)({\cal E}+2)^3{\cal E}^2=0  ~$$
has three real distinct roots out of six, namely the lowest 
$ {\cal E}_0(n=2)=-6$ with degeneracy $\delta=1$, 
the intermediate
$ {\cal E}_2(n=2)=-2$ with degeneracy $\delta=3$ 
and the highest one
$ {\cal E}_4(n=2)=0$ with degeneracy $\delta=2$. 

The corresponding orthogonal eigenvectors are 
\bea
&& \Psi_0(n=2)=  {\cal N}_0 
\left(\varphi^*_{1/2}\varphi^*_{3/2}+ \varphi^*_{1/2}\varphi^*_{5/2}
+ \varphi^*_{1/2}\varphi^*_{7/2} \right.
\nn\\
&&\left.\ \ \ \ \ \ \ \ +\varphi^*_{3/2}\varphi^*_{5/2}
+\varphi^*_{3/2}\varphi^*_{7/2}+ \varphi^*_{5/2}\varphi^*_{7/2}\right)
\label{psi0_n2}\\
&& \Psi_2(n=2)= {\cal N}_2 \left[
a_1 \left(\varphi^*_{1/2}\varphi^*_{3/2}-\varphi^*_{5/2}\varphi^*_{7/2}\right) 
 + a_2 \left(\varphi^*_{1/2}\varphi^*_{5/2}-
\varphi^*_{3/2}\varphi^*_{7/2}\right)
\right.\nn\\
&& \ \ \ \ \ \ \ \  \left. + a_3 \left(\varphi^*_{1/2}\varphi^*_{7/2}-
\varphi^*_{3/2}\varphi^*_{5/2}\right) \right]\label{psi2_n2}\\
&& \Psi_4(n=2)={\cal N}_4 \left[
b_2\left(\varphi^*_{1/2}-\varphi^*_{7/2}\right)\left(\varphi^*_{5/2}-
\varphi^*_{3/2}\right) 
\right. \nn\\
&& \ \ \ \  \ \ \ \  \left. + b_3 \left(\varphi^*_{1/2}-\varphi^*_{5/2}\right)
\left(\varphi^*_{3/2}-\varphi^*_{7/2}\right)\right]\label{psi4_n2}~,
\eea
${\cal N}_0$, ${\cal N}_2$, ${\cal N}_4$ being normalization factors
and $a_1,a_2,a_3,b_2$ and $b_3$ free parameters. 

The case just discussed asks for some comments. First, 
as for $n=1$, the matrix representing ${\cal H}_P$ can be written in 
${{4}\choose{2}}!$ equivalent ways (not all 
distinct), each one corresponding to a different labelling of the states,
but leading to the same determinant.

Furthermore in the above example the correspondence 
between the number of parameters that enter into a given wave function
and the degeneracy of the corresponding eigenvalue is again apparent. 
In general to an eigenvalue ${\cal E}_v(n)$ of given seniority $v$ correspond
\cite{Ring}
\be
\delta_v
={{\Omega}\choose{\frac{v}{2}}} - {{\Omega}\choose{\frac{v}{2}-1}} 
\label{degenerazione}
\ee
independent states constructed using the building blocks (\ref{phi})
(in our convention a binomial coefficient with a negative lower index 
vanishes).
The $\delta_v$ is commonly referred to as the seniority degeneracy. 

Finally to address the general case we exploit the two lessons learned from the previous example. The first one concerns 
the seniority degeneracy: its very existence proves that a further reduction 
of the dimension of the basis can be achieved. The second  
is that the structure of the symmetric matrix of dimension 
${{\Omega}\choose{n}}$ associated with ${\cal H}_P$ should be
\footnote{As already noticed in the $n$=2 case, 
all of the ${{\Omega}\choose{n}}!$   
orderings of the states are equivalent.} 

\be
\left(\begin{array}{ccc}
{\cal E} +n &  & 0\lor 1\\
&\ddots  & \\
0\lor 1 & & {\cal E} +n  
\end{array}
\right) \ ,
\label{matrice_generica}
\ee
where the symbol $0\lor 1$ indicates that the upper (lower)
triangle of the matrix is filled with zeros and ones.  
Indeed the matrix elements of ${\cal H}_P$ are one when the 
bra and ket differ by the quantum state of one 
(out of $n$) pair, otherwise they vanish. 
The diagonal matrix elements simply count the number of pairs. 

An elementary combinatorial analysis then shows 
that the number of ones in each row 
(column) of the matrix (\ref{matrice_generica}) is given by $n(\Omega -n)$.
Indeed a non-vanishing matrix element has the row specified by $n$ 
indices whereas, of the indices identifying the column, $n-1$ should be 
extracted from those fixing the row in all possible ways, which
amounts to $n$ possibilities. The missing index should then be selected from among the remaining $\Omega-n$ ones: hence the formula $n(\Omega - n)$ 
follows.  

To write down explicitly the matrix (\ref{matrice_generica}) it
is convenient to divide 
the set of the $\Omega$ even Grassmann variables, whose quantum numbers 
identify the levels where the $n$ pairs are to be placed, into two 
subsets, one with $\Omega - \nu$ and the other with $\nu$ elements 
(to be referred to as I and II, respectively) with 
$n\le \nu \le \Omega-n$. 

A priori each of these partitions is valid in the sense that it leads to a 
basis with a dimension lower than the one of (\ref{base}). On the other hand
the physics of the pairing hamiltonian is such that for a system of 
$n$ pairs $n+1$ eigenvalues should be expected, no matter what the 
value of $\Omega$. They of course correspond to the breaking of 0, 
1, 2 $\cdots n$ pairs. A partition leading to an $n+1$ dimensional basis, 
for any $\Omega$ and without degeneracy, is the one which has $\nu =n$. 

Indeed in this instance the ${{\Omega}\choose{n}}$ entries of 
each row and column of the matrix can then
be grouped into $n+1$ sets, the first one corresponding to the $n$ pairs placed
in the $\Omega-n$ levels of I, the remaining $n$ levels of II being empty
(see Fig.~1a).

\begin{figure}[ht]
\centerline{
\psfig{figure=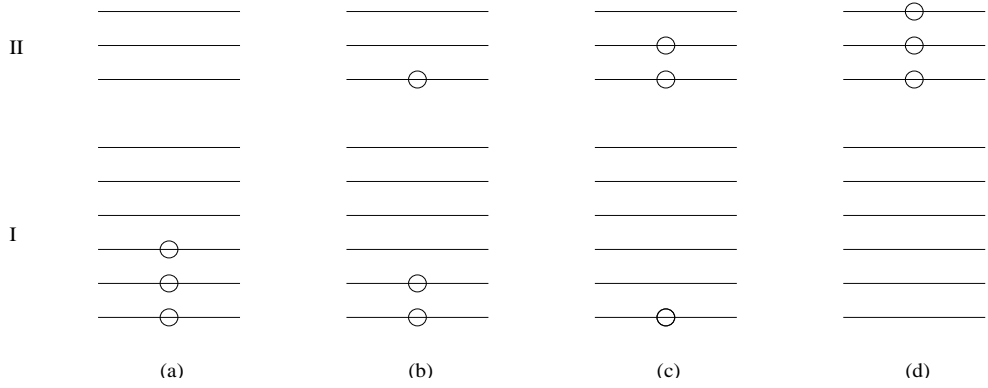,width=13cm,height=5cm,clip=}}
\caption[ ]{The figure shows, in the specific case $n$=3 and $\Omega$=9,
the partition of the $\Omega$ levels in the set I (with $\Omega -n$=6 levels)
and II (with $n=3$ levels).}

\end{figure}
 
To this first set correspond $d_0={{\Omega-n}\choose{n}}$ configurations. 
To the second set are associated configurations with $n-1$ pairs in I and 
one pair in II (see Fig.~1b), their number being 
$d_1=n {{\Omega-n}\choose{n-1}}$. In general the $(k+1)$-th set embodies 
configurations with $n-k$ pairs in I and $k$ pairs in II, their number being

\be
d_k = {{n}\choose{k}} {{\Omega-n}\choose{n-k}} \ \ \ \mbox{with}\ 
0\leq k\leq n\ \ (k\ \mbox{integer}).
\label{dk}
\ee
Clearly
$\sum_{k=0}^n d_k = {{\Omega}\choose{n}}$.

With this organization of the levels 
the matrix (\ref{matrice_generica}) splits into $(n+1)^2$ rectangular blocks 
$B_{kj}$ (with $0\leq k,j\leq n$) of dimension $d_k\times d_j$ and,
since the pairing Hamiltonian only connects states differing by the 
quantum number of one pair, the blocks with $|k-j|\geq 2$ will have 
vanishing elements. The matrix thus becomes {\it block-tridiagonal}. 

This suggests the 
introduction, {\it in lieu} of (\ref{phi}), of the following orthonormal set
of $n+1$ even commuting variables (the composite ``bosons'')

\be
\Phi_k^* = \frac{1}{\sqrt{d_k}} \sum_{m=1+s_{k-1}}^{s_k} 
\left[\varphi^*_{m_1}...\varphi^*_{m_n}\right]_m \ ,
\label{Phik}
\ee
where $k$ varies as in (\ref{dk}), $s_j = \sum_{l=0}^j d_l$, 
being $s_{-1}=0$ and $m$ again identifies the set $\{m_1, m_2 \cdots m_n\}$. 
Note that the states (\ref{Phik}), unlike those in (\ref{phi}), 
have in general an index of nilpotency higher than one. 

The definition (\ref{Phik}) also reflects our desire to have
the composite bosons retain as much as possible of the 
symmetry of $H_P$. And indeed the $\Phi^*_k$, while not fully 
symmetric with respect to the interchange of the $\varphi^*$, turn out 
to be invariant with respect to the 
interchange of the $\varphi^*$ belonging either to set I or to set II. 
This symmetry property stems from our choice (\ref{Phik}) which 
enforces the maximum coherence among the components of $\Phi^*_k$. 
It is remarkable that 
composite variables corresponding to different combinations of the 
$\varphi^*_{m_1} \varphi^*_{m_2} \cdots \varphi^*_{m_n}$ not only 
hold a lower symmetry than the one displayed by (\ref{Phik}), but may also lead to the wrong eigenvalues, as we have verified in some instances. 

Now in the minimal basis (\ref{Phik}) ${\cal H}_P$ is 
represented by a $(n+1)\times(n+1)$ matrix whose generic element 
$\left(M_n\right)_{ki}\equiv <\Phi_k^*|{\cal H}_P|\Phi_i^*>$ 
results from summing the elements of the
block $B_{ki}$ of the ${\Omega\choose n}\times{\Omega\choose n}$ matrix,
but for the normalization factor $1/\sqrt{d_k d_i}$.
The sum is performed by recognizing that all of the blocks 
have the same number of ones in each row.
Specifically, 
in the diagonal block $B_{kk}$ there are
$(n-k)(\Omega-2n+2k)$ ones in each row. In the upper diagonal block
$B_{k,k+1}$ each row instead contains 
\be
c_k=(n-k)^2
\label{ck}
\ee
ones. Since the total number of ones in each row of the matrix 
(\ref{matrice_generica}) is $n(\Omega-n)$, the number of ones in the 
rows of the lower diagonal block $B_{k,k-1}$ will be 
\be
b_k=k(\Omega-2n+k)\ . 
\label{bk}
\ee
As a consequence, the non-vanishing elements of the matrix $M_n$ turn out to be

\bea
\left(M_n\right)_{k,k+1} &=& <\Phi^*_k|{\cal H}_P|\Phi^*_{k+1}> = 
\sqrt{\frac{d_k}{d_{k+1}}} c_k~,
\\
\left(M_n\right)_{k+1,k} &=& <\Phi^*_{k+1}|{\cal H}_P|\Phi^*_k> =
\sqrt{\frac{d_{k+1}}{d_k}} b_{k+1}
\eea
and
\be
\left(M_n\right)_{kk} = <\Phi^*_k|{\cal H}_P|\Phi^*_k> = a_k \equiv 
{\cal E} + n + n(\Omega-n) - b_k - c_k\ .
\ee
Clearly $\left(M_n\right)_{k,k+1}=\left(M_n\right)_{k+1,k}$, 
since the operator ${\cal H}_P$ is Hermitian
and the basis (\ref{Phik}) is orthonormal.
The matrix thus becomes {\it tridiagonal}, reading

\bea
M_n = \left(
\begin{array}{ccccccc}
a_0 & \sqrt{\frac{d_0}{d_1}} c_0 &        0     & \cdot   & \cdot  & \cdot 
& \cdot\\
\sqrt{\frac{d_0}{d_1}} c_0 &  a_1 & \sqrt{\frac{d_1}{d_2}} c_1  &    0    
& \cdot  & \cdot & \cdot\\
   0    & \sqrt{\frac{d_1}{d_2}} c_1 &  a_2  & \sqrt{\frac{d_2}{d_3}} c_2 
&    0   & \cdot & \cdot\\
 \cdot  & \cdot   &     \cdot    & \cdot   & \cdot  & \cdot &\cdot\\   
 \cdot  &  0      &\sqrt{\frac{d_{k-1}}{d_k}}c_{k-1}& a_k 
&\sqrt{\frac{d_k}{d_{k+1}}} c_k & 0     & \cdot\\   
 \cdot  & \cdot   & \cdot        & \cdot   & \cdot  & \cdot &\cdot\\   
 \cdot  & \cdot   & \cdot        & \cdot   & 0      
&\sqrt{\frac{d_{n-1}}{d_n}}  c_{n-1} & a_n   
\end{array} \right)
\nonumber
\eea
and the associated eigenfunctions should be expanded in terms of 
the composite variables (\ref{Phik}), namely
\be
\psi (\Phi^*) = \sum_{k=0}^n u_k \Phi^*_k 
=  \sum_{k=0}^n \sqrt{d_k} w_k  \Phi^*_k
\ ,
\label{psi_Phi}
\ee
where the $\sqrt{d_k}$ is introduced for convenience.
The coefficients of the expansion 
(\ref{psi_Phi}) are then fixed by the eigenvalue equation

\be
M_n \vec u = 0 \ ,
\label{eqwk}
\ee
which, of the 
${\Omega\choose n}$ equations for the $\beta$'s,  reduces 
to $n+1$ equations for the $w$'s.  
The associated eigenvalues ${\cal E}$ obey the secular equation

\be
D_n = \det \left(M_n\right) = 0 \ ,
\label{DetD0}
\ee
where ${\cal E}$ enters into the diagonal matrix elements $a_k$.
 
Now from the general theory of symmetric tridiagonal matrices
one knows that (\ref{DetD0}) has $n+1$ 
{\em distinct} and {\em real} 
roots and that these are found by applying the recursive relation \cite{Pras}

\be
D_n = a_n D_{n-1} - \frac{d_{n-1}}{d_n} (c_{n-1})^2 D_{n-2}
\label{recD}
\ee
which, after some algebra, yields

\be
D_n = y(y-\Omega)[y-2(\Omega-1)][y-3(\Omega-2)] ... [y-n(\Omega-n+1)] 
\label{D_n}\ ,
\ee
where
\bea
y &=& n(\Omega-n)+{\cal E}+n \ .
\eea
We thus see that $D_n$ has all of the zeros of $D_{n-1}$ plus an extra one for
$y=n(\Omega-n+1)$.

Moreover (\ref{D_n}) allows us to write down for the general solution of 
(\ref{DetD0}) the expression 
\be
y=p(\Omega-p+1) \ \ \ \ \ \ \ \mbox{with} \ 0\le p \le n ~.
\ee
Hence {\it the well-known formula
for the spectrum of the pairing Hamiltonian}  

\be
{\cal E} = -(n-p)(\Omega-n-p+1) ~,
\ee
{\it is recovered},
the index $p=v/2$ being linked to the
seniority quantum number $v$.

For a tridiagonal matrix, a recursive relation among the eigenvectors 
components, similar to (\ref{recD}), can also be established.
In the specific case of the matrix $M_n$ it reads
\be
d_k c_k w_{k+1} = - d_{k-1} c_{k-1} w_{k-1} - d_k a_k w_k \ ,
\label{recw}
\ee
where again $0 \le k \le n$ and quantities with negative indices 
are meant to be zero. 

Thus for the lowest eigenvalue $y=0$ ($p=0$, zero seniority) we have

\be
w_0^{(v=0)}=w_1^{(v=0)}=....=w_n^{(v=0)} \ ,
\label{w0}
\ee
namely the collective state
\be
\psi_{v=0} ={\Omega\choose n}^{-1/2} \sum_{k=0}^n \sqrt{d_k} \Phi^*_k
= {\Omega\choose n}^{-1/2} \sum_{m=1}^{\Omega\choose n} 
\left[\varphi^*_{m_1}...\varphi^*_{m_n}\right]_m \ .
\label{sen0}
\ee
Indeed in (\ref{sen0})
all of the components of the wave function, i.e. the monomials
$\left[\varphi^*_{m_1}...\varphi^*_{m_n}\right]_m$, are coherently summed up.
This state obtains for a specific partition of the levels 
defining the matrix $M_n$. However, any other partition would lead to
the same result, the weights of the components all being equal. As a 
consequence the state (\ref{sen0}) has zero degeneracy.

Concerning the second eigenvalue $y=\Omega$ ($p=1$, seniority 2),
according to Eq.~(\ref{recw}) the components of its eigenstate are

\be
w_k^{(v=2)} = {\cal N}_2 \left(k\Omega - n^2\right) \ .
\label{w2}
\ee
 
We have found the following expression  
for the components of the state associated with a generic 
seniority $v$ 
\bea
w_k^{(v)}& = & {\cal N}_v \sum_{j=0}^{v/2} (-1)^j \frac{ {k \choose j} 
{v/2 \choose j} 
{\Omega -v/2 +1 \choose j}}{{n \choose j}^2}=\nn\\
&=& {\cal N}_v \ _3F_{2} \left( -k, -\Omega +\frac{v}{2} -1, -\frac{v}{2}; ~
-n, -n ;~ 1\right) 
\label{wk}
\eea
$ _3F_{2}$ being a generalized hypergeometric function. 
In the above the binomial $k \choose j$ is meant to vanish when 
$j > k$. 

In particular, for the vector of the maximum seniority $\vec w^{(v=2n)}$, 
corresponding to $y=n(\Omega-n+1)$ ($p=n$, seniority $2n$), 
one has 

\be
w_k^{(v=2n)} = {\cal N}_{2n} \left[(-1)^k 
\frac{(\Omega-2n+k)!(n-k)!}{(\Omega-2n)!}\right] \ .
\label{wn}
\ee

In (\ref{w2}), (\ref{wk}) and (\ref{wn}) ${\cal N}_2$, ${\cal N}_v$ and 
${\cal N}_{2n}$ are normalization constants. 

It is straightforward to show that the general eigenstates 
(\ref{psi_Phi}) reduce to the expressions (\ref{psi0_n2}, 
\ref{psi2_n2}, \ref{psi4_n2}) for $\Omega=4$ and $n=2$.

It is interesting in the present formalism to recover the following important result: 
for values of $\Omega$ large with respect to $n$, 
the number of dominant components in
$\vec w^{(v)}$ decreases with $v$, reflecting the weakening
of collectivity with increasing seniority.
Indeed in the limit $\Omega>>n$ the eigenvalues are

\be
y\simeq \frac{v}{2}\Omega 
\label{yv}
\ee
and moreover 
\be
c_k << \Omega \ ,
\ \ \ 
\frac{d_{k-1}}{d_k} c_{k-1} \simeq k\Omega
\ \ \ 
\mbox{and}
\ \ \ 
a_k \simeq \left(\frac{v}{2}-k\right)\Omega \ .
\ee
By inserting the above limits in (\ref{recw}) the $w_k^{(v)}$ 
components corresponding to the eigenvalues (\ref{yv}) are found to be 

\be
w_0^{(v)} = w_1^{(v)} = ... = w_{v/2-1}^{(v)} = 0
\ee
and
\be
w_k^{(v)} = \frac{{k\choose v/2}}{{n\choose v/2}} w_n^{(v)}
\ ,\ \ \ \ \ \ \ k=v/2,...,n \ .
\label{wkj}
\ee

We thus see that the state with seniority 
$v$ in the basis of the $\sqrt{d_k}\Phi^*_k$ and
in the large $\Omega$ limit has indeed $v/2$ vanishing components,
the remaining $n-v/2+1$ components being expressed through (\ref{wkj})
via the single component $w_n^{(v)}$.
In other words, when $\Omega$ is large the collectivity of a state 
with seniority 
$v $ decreases as $v$ increases, because its components become 
fewer and, furthermore, 
the surviving components are more capably expressed through a single one.

As a last point in this letter we briefly address the problem 
of the angular momentum of the 
composite bosons entering in our eigenstates. 
In this, for the sake of illustration we shall confine ourselves 
to the cases $n=1$ and $n=2$, and the latter only for $\Omega=4$. 
For this purpose it is natural to express the simple building blocks 
(\ref{phi}), 
the monomials, as superpositions of $\Omega$ composite bosons $B_J$ of 
{\it definite even}
\footnote{Since the $\varphi_m$ are even in $m$, only even $J$ are allowed.} 
{\it angular momentum with vanishing third component}, according to
\be
(-1)^{j-m}\varphi_m = \sqrt{\frac{2}{\Omega}}
\sum_J \langle jm,j-m | J0\rangle B_J~.
\label{phiPhi}
\ee
Moreover we account for the constraints the nilpotency of the $\varphi$'s
variables induces on the composite variables $B_J$. 
These constraints relate the latter among themselves, and hence the $B_J$ are no longer independent.

It then immediately follows that $B^*_0$,
a {\it composite s-boson},
coincides with the one pair wave function (\ref{psi0_n1}).
Similarly, with
a suitable choice of the parameters $\beta_m$,
the $v=2$ state (\ref{psi2_n1}) can be made to coincide with
$B^*_J$, with $J=2,4,\cdots 2(\Omega-1)$.
This state thus represents a broken pair. 

For two pairs in $\Omega=4$, the $v=0$ eigenstate (\ref{psi0_n2})
is given by the  product of two {\it composite s-bosons} (two unbroken pairs),
namely

\be
\Psi_0(n=2) = B_0^{*2}~.
\label{PHI0^2}
\ee

Likewise for the $v=2$ state (\ref{psi2_n2}) one has
\bea
&&\Psi_2(n=2) = 
\left(-4a_1-2a_2+a_3\right) B^*_0B^*_2 + \nn\\
&& + \left( 3a_1 -2a_2 +8a_3\right)\sqrt{\frac{3}{11}}  B^*_0 B^*_4 + 
\left(2a_1-5a_2-2a_3\right)\sqrt{\frac{7}{11}} B^*_0B^*_6\,,
\eea 
namely a wave function of good angular momentum $(J=2,4,6)$, 
as it is easily verified by appropriately choosing the parameters. 

Up to this point we have been able to express the eigenstates of $H_P$ only in terms of $s$ and $d$ composite bosons. This, however, is no longer the case for the $v=4$ eigenstate. Indeed, on the one 
hand it is remarkable that $B^*_0$ cancels out in the differences 
among the $\varphi^*$'s which enter into (\ref{psi4_n2}), an occurrence to
be expected since the state $v=4$ describes two broken pairs. 
On the other hand, from its explicit expression in term of $B_J$, namely
\bea
\lefteqn{\Psi_4(n=2) = \nn 
\frac{1}{7}\left( 5 b_3 -4 b_2\right) B_2^{*2}  - 
\frac{5}{77} \left( 11 b_3 + b_2\right) B_4^{*2}   +
\frac{7}{11} b_2 B_6^{*2}  +}\\
&&\!\!\!\!\!\!\!\!\!\!
-\frac{8}{7\sqrt{33}} \left( 5 b_3 -4 b_2\right) B_2^* B_4^*  -
\frac{4}{\sqrt{77}} \left(b_3 -3 b_2\right) B_2^* B_6^*  +
\frac{4}{11\sqrt{21}} \left( 11 b_3 +2 b_2\right) B_4^* B_6^* 
\nn\\
\label{v4Phi}
\eea  
it appears that (\ref{v4Phi}) cannot be given only in terms 
of $s$ and $d$ composite bosons.
Note that the wave functions discussed here are yet to be normalized. 

It is worth pointing out that, 
since our basis is clearly incomplete insofar as the angular momentum is concerned, in general we are bound to obtain wave functions that contain superpositions of differing angular momenta. 

In conclusion the aim of this letter has been to illustrate for the specific 
example of the pairing hamiltonian, viewed as a prototype for the interaction 
in complex many-body systems, the suitability of the Grassmann algebra 
for dealing with composite fields. This approach permits a major reduction in the number of 
degrees of freedom used to describe many-body systems, in particular to treat the discrete states of nuclei. 

The relevant outcomes of the present analysis arise from the building blocks used in setting up our minimal basis. These, in fact, 
\begin{itemize}
\item
keep track of the nature of their constituents through a finite index of nilpotency;
\item
while not fully symmetric in the exchange of the $\varphi$'s, nevertheless fulfill a remarkable symmetry property. 
\end{itemize}

Finally, it is seen that more than two composite bosons are required to treat the dynamics of $H_P$.

\section*{Acknowledgements} The authors like to thank Dr. M. Caselle for many 
useful discussions and Prof. T.W. Donnelly for critical remarks and 
constructive comments.


\begin{thebibliography}{99}
\bibitem{Iac}
F.~Iachello and A.~Arima, 
The interacting boson model,
(Cambridge University Press, Cambridge, 1897) 

\bibitem{Bar97}
M.B. Barbaro, A. Molinari and F. Palumbo, 
Nucl. Phys. B 487 (1997) 492

\bibitem{Pal98}
S.~Caracciolo and F.~Palumbo, 
Nucl. Phys. B 512 (1998) 505

\bibitem{Pal99}
F.~Palumbo, 
Phys. Rev. D 60 (1999) 074009

\bibitem{Fad80}
L.D. Faddeev and A.A. Slavnov, 
Gauge Fields: Introduction to Quantum Theory (Benjamin, 1980)

\bibitem{Ring}
P.~Ring and P.~Shuck, 
The Nuclear Many Body Problem,
(Springer-Verlag, New York, 1980), p.~225 

\bibitem{Pras}
V.V. Prasolov, 
Problems and Theorems in Linear Algebra, 
(American Mathematical Society, Providence, 1994), p.~65

\end{thebibliography}
\end{document}